\documentclass[%
preprint, 
 amsmath,amssymb,
 aps, physrev
]{revtex4-2}

\usepackage{graphicx}% Include figure files
\usepackage{dcolumn}% Align table columns on decimal point
\usepackage{bm}% bold math

\usepackage[utf8]{inputenc} % Unicode input encoding
\usepackage[T1]{fontenc} % Font encoding
\usepackage{lmodern} % Use an updated latim modern font (same as latex, just updated)
\usepackage{siunitx}

\begin{document}

%\preprint{APS/123-QED}

\title{\textbf{Kelvin‐Wave-Inspired Optical Vortex Excitation in Kerr Nonlinear Media} 
}% 

\author{Yosuke Minowa}

 \email{Contact author: minowa.yosuke.7m@kyoto-u.ac.jp}
\affiliation{%
Department of Physics, Kyoto University, Kitashirakawa-Oiwake-cho, Sakyo-ku, Kyoto, Japan
}%
\affiliation{The Hakubi Center for Advanced Research, Kyoto University, Kitashirakawa-Oiwake-cho, Sakyo-ku, Kyoto, Japan}

\author{Nobuhiko Yokoshi}
\affiliation{
Department of Physics and Electronics, Osaka Metropolitan University, Osaka 599-8531, Japan
}%

\author{Makoto Tsubota}
\affiliation{%
Department of Physics, Nambu Yoichiro Institute of Theoretical and Experimental Physics (NITEP), Osaka Metropolitan University, 3-3-138 Sugimoto, 558-8585 Osaka, Japan
}%

\date{\today}% It is always \today, today,

\begin{abstract}
We demonstrate a direct one-to-one correspondence between nonlinear optical fields in defocusing Kerr media and wave functions in weakly interacting Bose–Einstein condensates or quantum fluids. Based on this correspondence, we propose the existence of excitations in an optical vortex beam characterized by a helical deformation of its phase singularity core. These excitations are direct analogues of Kelvin waves known in quantum and classical fluid dynamics. We further show that the excitations exhibit two distinct branches, one of which includes a stationary solution. A feasible experimental scheme for generating these excitations is also discussed.
\end{abstract}

\maketitle

%\tableofcontents

\section{Introduction}
As beautifully illustrated in Leonardo da Vinci’s sketches (e.g., the Royal Collection at Windsor, RCIN 912660v), vortices have fascinated humans by appearing in various forms and scales. From ordinary whirlpools and draining water vortices to quantized vortices, our world is abundant with vortex structures. On an even broader sense, vortex-like entities include flux quanta in superconductors, magnetic skyrmions, and acoustic vortices, among others. Optical vortex beams have attracted considerable attention in recent decades due to their unique property of carrying orbital angular momentum\cite{allenOrbitalAngularMomentum1992}, resulting in chiral light-matter interactions\cite{wongExcitationOrbitalAngular2012, hashiyadaRapidModulationLeft2024,taguchiNanoscaleChiralityEnhancement2025} capable of driving nano/micromechanical rotors\cite{heDirectObservationTransfer1995,simpsonMechanicalEquivalenceSpin1997}, nanofabrication\cite{toyodaTransferLightHelicity2013,yuyamaFabricationArrayHemispherical2023,matsumotoHelicalSurfaceRelief2025}, and chiral crystallization\cite{toyodaChiralCrystallizationManipulated2023}.

Similarities between fluid vortices and optical vortices have been recognized and discussed, especially in the context of spatial optical solitons in nonlinear optical media\cite{kivsharDynamicsOpticalVortex1998,kivsharOpticalSolitonsFibers2003,pelinovskySelffocusingPlaneDark1995,swartzlanderOpticalVortexSolitons1992}. Optical vortex solitons\cite{snyderStableBlackSelfguided1992} are often considered analogues of quantized vortices\cite{pitaevskiiVortexLinesImperfect1961}, and their behaviors have been studied using quantum fluid dynamics analogies\cite{rozasPropagationDynamicsOptical1997,kivsharDynamicsOpticalVortex1998}. Although their similarity is qualitatively evident, it is not yet settled to what extent the relationship constitutes a rigorous correspondence. This uncertainty arises mainly due to the fundamental difference in governing equations for nonlinear optical waves and quantum fluids. The nonlinear Schrödinger equation, widely used to describe optical solitons in Kerr media\cite{boydNonlinearOpticsThird2008}, is a (2+1)-dimensional equation, with two transverse spatial dimensions and one longitudinal coordinate (beam propagation direction) serving as an evolution parameter analogous to time. By contrast, the Gross–Pitaevskii equation for quantum fluids includes both temporal and three spatial degrees of freedom\cite{donnellyQuantizedVorticesHelium1991}. Efforts to reduce the Gross–Pitaevskii equation from three spatial dimensions to two or one\cite{kivsharOpticalSolitonsFibers2003,rozasPropagationDynamicsOptical1997} obscure the precise correspondence of three-dimensional dynamics between the two systems, such as deformation or vortex-core motion and excitations on vortices. Although the concept of spatiotemporal solitons---often referred to as ``light bullets'' \cite{desyatnikovChapter5Optical2005}---deals with three spatial dimensions, it still treats the propagation direction differently, and its relation to the Gross-Pitaevskii equation is not evident.

\begin{figure}
\includegraphics[width=0.45\textwidth]{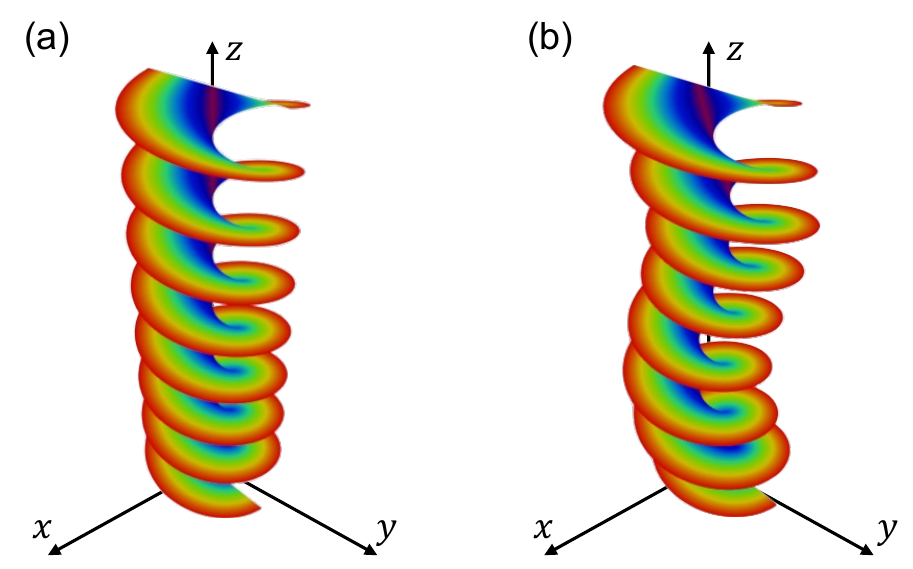}% Here is how to import EPS art
\caption{\label{fig:3d}Schematic illustration of optical vortex beams. (a) Phase front of a straight optical vortex beam. (b) Phase front of an optical vortex beam exhibiting a helically deformed phase singularity, analogous to a Kelvin-wave excitation observed on quantized vortices. The entire electric field distribution is helically deformed.}
\end{figure}

In this paper, we present a direct correspondence between optical vortex beams in defocusing Kerr media and quantized vortices in quantum fluids. We demonstrate that optical vortex dynamics can be described by a (3+1)-dimensional nonlinear Schrödinger equation, formally identical to the Gross–Pitaevskii equation. As an illustrative example, we show that a helical deformation of the phase singularity in optical vortex beams can occur as schematically shown in Fig. \ref{fig:3d}. This deformation corresponds to the well-known Kelvin wave\cite{barenghiQuantumTurbulence2023,minowaDirectExcitationKelvin2025}, a fundamental excitation on quantized vortices\cite{saffmanVortexDynamics1993}. We discuss the possible excitation schemes to generate the Kelvin-wave analogue modes in nonlinear optics.

\section{(3+1)-dimensional nonlinear Schrödinger equation in Kerr nonlinar media}
Here, we derive a nonlinear Schrödinger equation describing the spatiotemporal dynamics of nonlinear optical waves in Kerr media. The resulting equation has the same form as the Gross–Pitaevskii equation governing weakly interacting Bose–Einstein condensates or quantum fluids~\cite{donnellyQuantizedVorticesHelium1991,barenghiQuantumTurbulence2023}. The nonlinear Schrödinger equation commonly employed in nonlinear optics for third-order nonlinearities~\cite{kivsharDarkSolitonsBackgrounds1994,kivsharDynamicsOpticalVortex1998,swartzlanderOpticalVortexSolitons1992,desyatnikovChapter5Optical2005} takes the form:
\begin{equation}
    i\frac{\partial A}{\partial z} + \frac{1}{2k_0}\nabla_{\perp}^{2}A + \frac{\omega_0^2}{2k_0\,c_m^2}\,\chi^{(3)}\,|A|^2 A = 0,
\end{equation}
where $A$ is the slowly varying envelope of the wave, $\nabla_{\perp}$ is the transverse gradient, and $\chi^{(3)}$ is the third-order nonlinear susceptibility. Parameters $k_0$, $\omega_0$, and $c_m = c/n_0$ are the wavenumber, angular frequency, and the speed of light in the medium, respectively, satisfying the linear dispersion relation $\omega_0 = c_m k_0$. Note that this dispersion relation accounts only for the linear refractive index $n_0$. This widely used form is derived under the assumptions of slowly varying envelopes and the paraxial approximation. We will derive a more general form starting from the nonlinear wave equation without requiring the paraxial approximation~\cite{boydNonlinearOpticsThird2008}:
\begin{equation}
    \nabla^{2}\mathbf{E} - \frac{1}{c_m^{2}}\frac{\partial^{2}\mathbf{E}}{\partial t^{2}} = \mu_{0}\frac{\partial^{2}\mathbf{P}_{\text{NL}}}{\partial t^{2}},\label{eq:nonlwave}
\end{equation}
where $\mathbf{P}_{\text{NL}}$ denotes the nonlinear polarization. Apart from the linear polarization contribution in $c_m$, we focus exclusively on the third-order nonlinear term defined as
\begin{equation}
    \mathbf{P}_{\text{NL}} = \varepsilon_{0}\varepsilon\,\chi^{(3)}\,|\mathbf{E}|^{2}\,\mathbf{E}.
\end{equation}
For clarity, we have factored out the linear dielectric constant $\varepsilon=n_0^2$ from $\chi^{(3)}$. Assuming an isotropic material and considering a scalar equation, we have
\begin{equation}
    \nabla^{2}E - \frac{1}{c_m^{2}}\frac{\partial^{2}E}{\partial t^{2}} = \mu_{0}\frac{\partial^{2}}{\partial t^{2}}\left[\varepsilon_{0}\varepsilon\,\chi^{(3)}\,|E|^{2}\,E\right].\label{eq:nonlinear_wave}
\end{equation}
We simplify the equation by introducing the slowly varying envelope:
\begin{equation}
    E(\mathbf{r},t) = \Psi(\mathbf{r},t)\exp\left[ik_0 z - i\omega_{0}t\right].
\end{equation}
Expanding the second-order time derivative in Eq.~(\ref{eq:nonlinear_wave}), we obtain
\begin{equation}
\begin{aligned}
    \frac{\partial^{2}E}{\partial t^{2}} 
    &= \frac{\partial^{2}}{\partial t^{2}}\left[\Psi e^{ik_0z-i\omega_{0}t}\right] \\
    &= e^{ik_0z-i\omega_{0}t}\left(\frac{\partial^{2}\Psi}{\partial t^{2}} - 2i\omega_{0}\frac{\partial\Psi}{\partial t} - \omega_{0}^{2}\Psi\right).
\end{aligned}
\end{equation}
Neglecting the second-order derivative of the slowly varying envelope, we write
\begin{equation}
    \frac{\partial^{2}E}{\partial t^{2}} \approx e^{ik_0z-i\omega_{0}t}\left(-2i\omega_{0}\dfrac{\partial\Psi}{\partial t} - \omega_{0}^{2}\Psi\right).
\end{equation}
The nonlinear polarization,
\begin{equation}
    P_{\mathrm{NL}} = \varepsilon_{0}\varepsilon\chi^{(3)}|\Psi|^{2}\Psi\, e^{ik_0z-i\omega_{0}t},
\end{equation}
is already small, thus retaining only the lowest-order contribution in Eq.~(\ref{eq:nonlinear_wave}), we have
\begin{equation}
\begin{aligned}
    \mu_{0}\frac{\partial^{2}P_{\mathrm{NL}}}{\partial t^{2}}
    &\approx -\mu_{0}\varepsilon_{0}\varepsilon\,\omega_{0}^{2}\chi^{(3)}|\Psi|^{2}\Psi\, e^{ik_0z-i\omega_{0}t}\\
    &= -\frac{\omega_{0}^{2}}{c_m^{2}}\chi^{(3)}|\Psi|^{2}\Psi\, e^{ik_0z-i\omega_{0}t}.
\end{aligned}
\end{equation}
Next, the spatial derivative of the electric field expands as
\begin{equation}
    \nabla^{2}E = e^{i(k_{0}z-\omega_{0}t)}\left[\nabla^{2}\Psi + 2ik_{0}\dfrac{\partial\Psi}{\partial z} - k_{0}^{2}\Psi \right].
\end{equation}
Combining all terms, we arrive at the equation:
\begin{equation}
    \nabla^{2}\Psi + 2ik_{0}\dfrac{\partial\Psi}{\partial z} + 2i\frac{\omega_{0}}{c_m^{2}}\dfrac{\partial\Psi}{\partial t} = -\frac{\omega_0^2}{c_m^2}\chi^{(3)}|\Psi|^{2}\Psi.
\end{equation}
We now introduce new variables\cite{sprangleNonlinearInteractionIntense1990}, defined by
\begin{equation}
    \zeta = z - c_m t,\quad \tau = t.
\end{equation}
The partial time derivative transforms as
\begin{equation}
    \frac{\partial}{\partial t}\Big|_{z} = \frac{\partial}{\partial \tau} - c_m\frac{\partial}{\partial \zeta}.
\end{equation}
Thus, we obtain the nonlinear Schrödinger equation for third-order nonlinear materials:
\begin{equation}
\begin{aligned}
    i\dfrac{\partial\Psi}{\partial \tau} &= -\frac{c_m^2}{2\omega_0}\nabla'^2\Psi - \frac{\omega_0}{2}\chi^{(3)}|\Psi|^{2}\Psi \\
    &= -\frac{c_m^2}{2\omega_0}\nabla'^2\Psi + g|\Psi|^{2}\Psi.\label{eq:gplight}
\end{aligned}
\end{equation}
where $\nabla'$ is the gradient operator including $\zeta$.
This equation corresponds directly to the Gross–Pitaevskii equation for quantum fluids~\cite{fetterVorticesTrappedDilute2001},
\begin{equation}
    i\hbar\frac{\partial\Psi}{\partial t} = -\frac{\hbar^2}{2m}\nabla^2\Psi + g_q|\Psi|^2\Psi.\label{eq:gpquantum}
\end{equation}
The defocusing Kerr media ($\chi^{(3)}<0$) corresponds to a positive $g_q$ value for normal repulsive Bose–Einstein condensates. This analogy aligns with the stability of optical vortex beams in defocusing Kerr media, contrasting with their instability in focusing Kerr media~\cite{desyatnikovChapter5Optical2005}. Note that the argument from Eq. (\ref{eq:nonlwave}) to Eq. (\ref{eq:gplight}) is actually parallel to the derivation of the Schrödinger equation as the non-relativistic limit of the Klein–Gordon equation\cite{zeeQuantumFieldTheory2010}.

\section{Kelvin wave analogue excitation in defocusing Kerr media}
As an intriguing application of the correspondence relation derived above, we now demonstrate the existence of an optical vortex beam solution possessing a helically deformed phase singularity. This solution is analogous to the well-known Kelvin wave observed along quantized and classical vortices~\cite{thomsonVibrationsColumnarVortex1880,minowaDirectExcitationKelvin2025,fondaDirectObservationKelvin2014,vinenKelvinWaveCascadeVortex2003,kivotidesKelvinWavesCascade2001}. The mode structure and dispersion relation of Kelvin waves in quantum fluids have been derived from the Gross–Pitaevskii equation through various approaches~\cite{pitaevskiiVortexLinesImperfect1961,rowlandsVibrationsQuantizedVortex1973,robertsVortexWavesCompressible2003}. Here, we employ a heuristic ansatz method~\cite{kobayashiKelvinModesNambu2014a}.

We start from the straight vortex solution of Eq. (\ref{eq:gplight}). Note that solving this equation requires a finite background in the far field. Therefore, in realistic scenarios, a finite volume (such as a waveguiding structure) is necessary, analogous to the situation of trapped Bose–Einstein condensates~\cite{fetterVorticesTrappedDilute2001} or superfluid helium in containers~\cite{tangImagingQuantizedVortex2023,makinenRotatingQuantumWave2023,perettiDirectVisualizationQuantum2023,minowaVisualizationQuantizedVortex2022}. Throughout our discussion, we assume a large but finite system size ($r=R$) which is determined by beam waist here and non-zero boundary conditions in the farfield.

Then, the straight vortex beam ansatz can be expressed as~\cite{donnellyQuantizedVorticesHelium1991}
\begin{equation}
    \Psi = C f(r) e^{i l \theta} e^{-i \mu \tau},
\end{equation}
where $r$ and $\theta$ are cylindrical coordinates, and $l$ is an integer indicating the topological charge. We consider only mode with $|l|= 1$ since they are known to be stable~\cite{desyatnikovChapter5Optical2005}. The parameter $\mu$ is chosen to satisfy the far-field boundary condition through the coefficient $C$ (see below). By substituting this ansatz into Eq.~(\ref{eq:gplight}), we obtain a differential equation for $f(r)$:
\begin{equation}
    f = -\frac{c_m^2}{2\omega_0\mu}\left[\frac{1}{r}\frac{\partial}{\partial r}\left(r\frac{\partial f}{\partial r}\right)-\frac{1}{r^2}f\right] + \frac{g C^2}{\mu} f^3.
\end{equation}

Introducing the normalized variable
\begin{equation}
    \xi = \frac{r}{r_0} = \frac{r}{c_m/\sqrt{2\omega_0\mu}}
\end{equation}
and choosing the constant factor $C$ such that
\begin{equation}
    \frac{g C^2}{\mu} = 1,
\end{equation}
the resulting dimensionless equation for $f(\xi)$ becomes
\begin{equation}
    f(\xi) = -\left[\frac{1}{\xi}\frac{\partial}{\partial \xi}\left(\xi\frac{\partial f}{\partial \xi}\right)-\frac{1}{\xi^2}f\right] + f^3.
\end{equation}
The solution to this equation has been extensively studied\cite{pitaevskiiVortexLinesImperfect1961,donnellyQuantizedVorticesHelium1991}, and its asymptotic behavior is known to be $f \propto \xi$ as $\xi \rightarrow 0$.

The Lagrangian density corresponding to Eq.~(\ref{eq:gplight}) is given by~\cite{kobayashiKelvinModesNambu2014a}
\begin{equation}
    \mathcal{L} = \frac{i}{2}\left(\Psi^*\dfrac{\partial\Psi}{\partial \tau}- \Psi\dfrac{\partial\Psi^*}{\partial \tau}\right) - \frac{c_m^2}{2\omega_0}|\nabla' \Psi|^2 - \frac{1}{2}g|\Psi|^4.\label{eq:ldensity}
\end{equation}
Now, we introduce a slightly deformed vortex ansatz of the form
\begin{equation}
    \Psi = C f(\bar{r}) e^{i l \bar{\theta}-i \mu \tau},
\end{equation}
where $\bar{r}$ and $\bar{\theta}$ incorporate the deformation through the vortex core position $X = X(\zeta, \tau)$ and $Y = Y(\zeta, \tau)$, as follows:
\begin{align}
    \bar{r} &= \sqrt{(x - X)^2 + (y - Y)^2},\\
    \bar{\theta} &= \tan^{-1}\left(\frac{y - Y}{x - X}\right).
\end{align}

After integrating over the $xy$-plane, we obtain the effective Lagrangian:
\begin{equation}
\begin{aligned}
    L_{\mathrm{eff}} =&\, l C^2 \pi (Y\dot{X} - X\dot{Y}) \\
    &- \frac{\pi c_m^2 l^2 C^2}{2\omega_0}(Y_\zeta^2 + X_\zeta^2)\ln\left(\frac{R}{r_0}\right).
\end{aligned}
\end{equation}

Using the Euler–Lagrange equations for $X$ and $Y$, namely,
\begin{equation}
    \frac{\partial L_{\mathrm{eff}}}{\partial X} - \partial_\tau\left(\frac{\partial L_{\mathrm{eff}}}{\partial \dot{X}}\right) - \partial_\zeta\left(\frac{\partial L_{\mathrm{eff}}}{\partial X_\zeta}\right) = 0,
\end{equation}
and similarly for $Y$, we derive the equations of motion for $X$ and $Y$:
\begin{equation}
\begin{aligned}
    -\dot{Y} + \frac{l c_m^2}{2\omega_0}X_{\zeta\zeta}\ln\left(\frac{R}{r_0}\right) &= 0,\\[6pt]
    \dot{X} + \frac{l c_m^2}{2\omega_0}Y_{\zeta\zeta}\ln\left(\frac{R}{r_0}\right) &= 0.\label{eq:eom}
\end{aligned}
\end{equation}

The derived coupled equations (\ref{eq:eom}) admit natural helical solutions such as
\begin{equation}
\begin{aligned}
    X_R &= \delta\cos(K_R \zeta \mp \Omega_R \tau),\\
    Y_R &= \delta\sin(K_R \zeta \mp \Omega_R \tau),
\end{aligned}
\end{equation}
for right-handed helical deformations, and
\begin{equation}
\begin{aligned}
    X_L &= \delta\cos(-K_L \zeta \pm \Omega_L \tau),\\
    Y_L &= \delta\sin(-K_L \zeta \pm \Omega_L \tau),
\end{aligned}
\end{equation}
for left-handed helical deformations, where $\delta > 0$ is the amplitude. These solutions directly correspond to the Kelvin-wave modes in quantum and classical fluids~\cite{pitaevskiiVortexLinesImperfect1961,thomsonVibrationsColumnarVortex1880}. Here, the upper (lower) sign denotes waves propagating in the positive (negative) $\zeta$ direction. Note that the propagation direction is determined by both the handedness of the Kelvin waves and the sign of $l$ ~\cite{donnellyQuantizedVorticesHelium1991,minowaDirectExcitationKelvin2025}. Substituting these solutions into the equations of motion, we obtain the well-known dispersion relations:
\begin{equation}
\begin{aligned}
    \pm\Omega_R &= K_R^2 \frac{l c_m^2}{2\omega_0}\ln\left(\frac{R}{r_0}\right),\\[6pt]
    \mp\Omega_L &= K_L^2 \frac{l c_m^2}{2\omega_0}\ln\left(\frac{R}{r_0}\right).
\end{aligned}
\end{equation}

Next, transforming back from the shifted variable $\zeta$ and $\tau$, we have
\begin{equation}
\begin{aligned}
    X_R &= \delta\cos(K z - K c_m t \mp \Omega t)\\
        &= \delta\cos\bigl(K z \mp (\Omega \pm K c_m) t\bigr),\\[6pt]
    X_L &= \delta\cos(-K z + K c_m t \pm \Omega t)\\
        &= \delta\cos\bigl(-K z \pm (\Omega \pm K c_m) t\bigr).
\end{aligned}
\end{equation}

\begin{figure}
    \centering
    \includegraphics[width=0.45\textwidth]{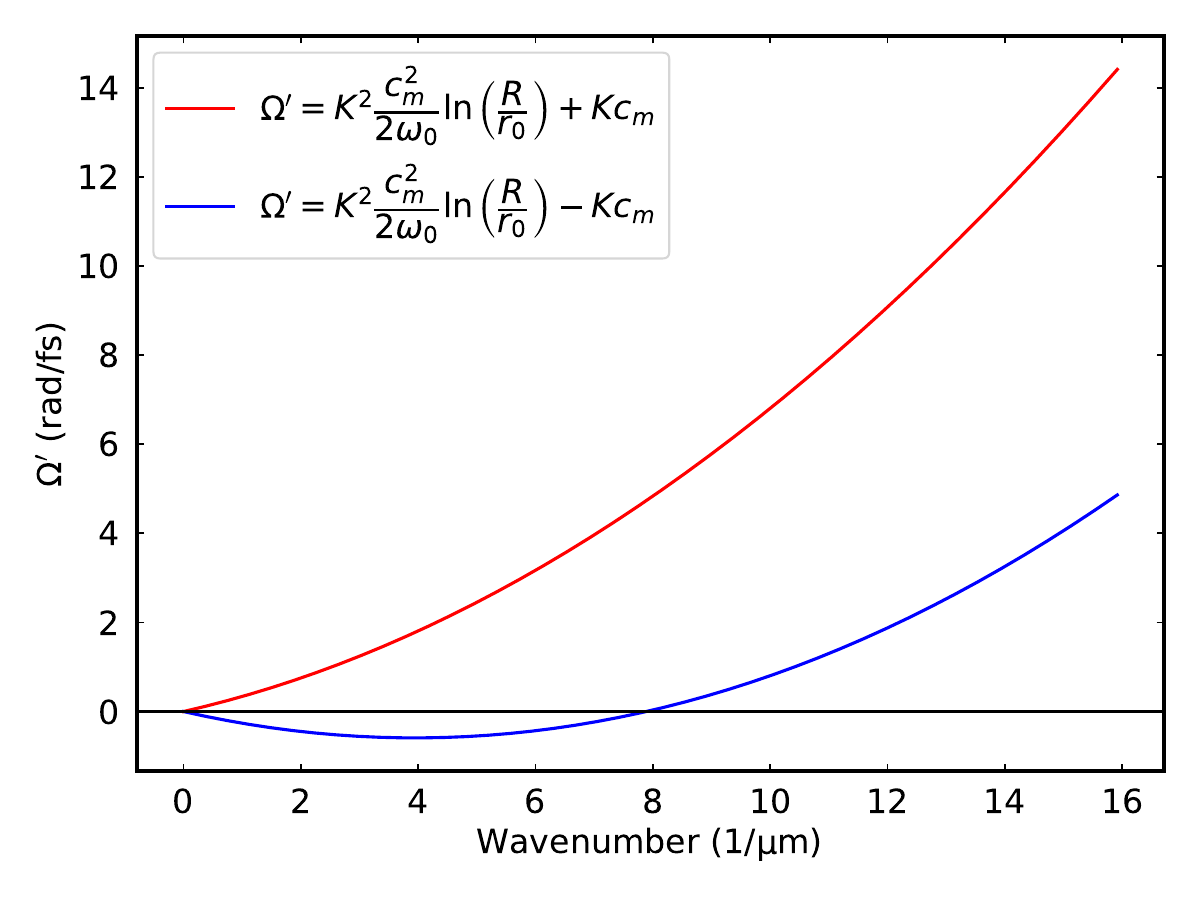}
    \caption{\label{fig:dispersion} Dispersion relation for the Kelvin-wave analogue mode along the vortex beam. Here we assume $\ln\left(R/r_0\right)=5$.}
\end{figure}

Thus, the dispersion relations become
\begin{equation}
\begin{aligned}
    \Omega_R' &= \Omega_R \pm K_R c_m\\
              &= \pm K_R^2\frac{l c_m^2}{2\omega_0}\ln\left(\frac{R}{r_0}\right) \pm K_R c_m,\\[6pt]
    \Omega_L' &= \Omega_L \pm K_L c_m\\
              &= \mp K_L^2\frac{l c_m^2}{2\omega_0}\ln\left(\frac{R}{r_0}\right) \pm K_L c_m.\label{eq:dispersion}
\end{aligned}
\end{equation}

Interestingly, we identify a stationary solution $\Omega_{R,L}' = 0$, where the helical deformation of the vortex beam core remains stationary. We note that the stationary Kelvin-wave solutions—namely, the right-handed wave on the $l=+1$ vortex and the left-handed wave on the $l=-1$ vortex—are known to propagate in the negative $\zeta$ direction~\cite{donnellyQuantizedVorticesHelium1991,minowaDirectExcitationKelvin2025}. Therefore, we understand that Kelvin waves propagating opposite to the direction of light result in stationary solutions.

To illustrate a typical stationary solution, we plot the dispersion relation (\ref{eq:dispersion}) for $\ln\left(R/r_0\right)=5$ in Fig.~\ref{fig:dispersion}, setting $n_0=1$ and $\lambda=2\pi/k_0=532\,\mathrm{nm}$ for simplicity. Two branches of the dispersion relation appear, with one branch exhibiting a zero-crossing point, representing the stationary Kelvin-wave mode. Both branches follow a squared wavenumber dependence as $K\rightarrow\infty$.

To further illustrate the stationary solution, we consider the spatial profile of the Kelvin-wave mode. The electric field of the Kelvin-wave analogue excitation is expressed as
\begin{equation}
\begin{aligned}
    E(\mathbf{r},t)
    &= C f(\bar{r})\, e^{i l\bar{\theta}-i\mu t}\,\exp\bigl(i k_0 z - i \omega_0 t\bigr)\\[4pt]
    &= E_0 f(\bar{r})\, e^{i \bar{\theta}}\exp\bigl(i k_0 z\bigr).
\end{aligned}
\end{equation}

\begin{figure*}
\includegraphics[width=1.0\textwidth]{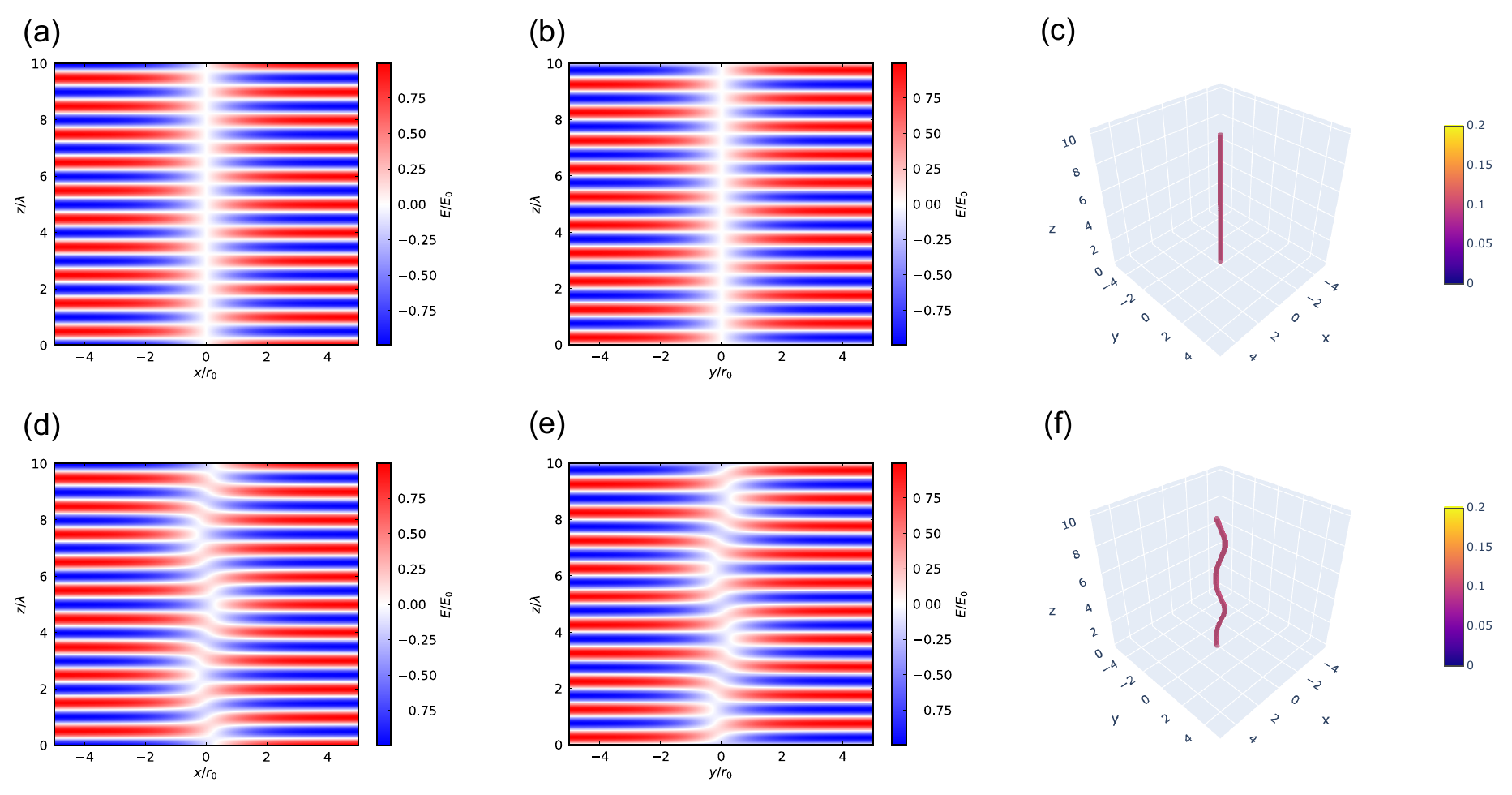}% Here is how to import EPS art
\caption{\label{fig:distribution}Electric field distribution around the vortex core for the straight vortex solution (a--c) and the left-handed stationary Kelvin-wave solution (d--f). Panels (a, d) and (b, e) illustrate the field distribution in the $xz$ and $yz$ planes, respectively. Panels (c, f) show the iso-surface of the normalized electric field magnitude, $|E/E_0|$.}
\end{figure*}

We present the electric-field distributions alongside the vortex beam core structure around the vortex core region, comparing the straight vortex solution (Fig.~\ref{fig:distribution}(a)-(c)) with the left-handed stationary Kelvin-wave solution (Fig.~\ref{fig:distribution}(d)-(f)). We use the same parameters as in Fig.~\ref{fig:dispersion}, with $\delta=0.3$ and an approximate profile function $f(\xi)\approx\tanh(0.582\,\xi)$~\cite{pethickBoseEinsteinCondensation2008,ishimoriDynamicsTopologicalVortices1986}. Figures~\ref{fig:distribution}(a,d) and (b,e) depict the electric field in the $xz$ and $yz$ planes, respectively, while Figures~\ref{fig:distribution}(c,f) illustrate the three-dimensional vortex-core structure using isosurface plots for $0\leq|E/E_0|\leq0.2$. These results clearly highlight the helical nature of the stationary Kelvin-wave mode.

\section{Discussion}
Experimental observation of stationary or non‐stationary Kelvin‐wave analog excitations is possible either by directly monitoring the transverse shift of the vortex beam or by imaging its path through a weakly scattering nonlinear medium from the side. A straightforward method to observe the helical deformation of the optical vortex beam is to measure the shift in its intensity pattern after passing through the nonlinear medium. Kelvin‐wave analog modes can be excited by periodically modulating the optical properties of the medium, in accordance with the recent experimental demonstration of Kelvin‐wave excitation in superfluid helium\cite{minowaDirectExcitationKelvin2025}. The modulation can be achieved via an additional pump beam or through electro‐optic or magneto‐optic effects. Direct imaging of the deformed beam path is also feasible. As an example, we consider the stationary solution. Using Eq. (\ref{eq:dispersion}) and setting $\Omega'=0$, we obtain $2\Lambda/\lambda=\ln(R/r_0)=\ln(\frac{\omega_0\sqrt{\chi^{(3)}}}{c_m}\sqrt{\frac{2P}{\pi c n_0\epsilon_0}})$, where $P$ is the optical beam power and $\Lambda$ and $\lambda$ are the wavelengths of the Kelvin wave and the optical wave, respectively. Assuming a 1 W beam propagating through a rubidium vapor cell with a nonlinear refractive index of $10^{-6}\,\mathrm{cm}^2/\mathrm{W}$\cite{wangMeasurementKerrNonlinear2020}, we obtain $2\Lambda/\lambda \sim 3.5$, which ensures that direct imaging of the stationary deformed shape of the optical beam path is experimentally feasible.

Finally, the direct correspondence demonstrated here between optical and quantized vortices provides valuable insights for both optics and quantum fluid dynamics. On one hand, quantized vortex dynamics are expected to have direct analogues in nonlinear optics. One important example is the possible reconnection of two optical vortex beams\cite{minowaVisualizationQuantizedVortex2022}, which facilitates efficient interactions between optical vortices and may enable the exchange of information encoded in the beams\cite{bozinovicTerabitScaleOrbitalAngular2013}. On the other hand, an optical simulator for quantum fluid dynamics may be realized. Phenomena involving multiple quantized vortices, such as Kelvin-wave cascades\cite{kivotidesKelvinWavesCascade2001,vinenKelvinWaveCascadeVortex2003}, remain both experimentally and computationally challenging. Multiple optical vortex beams in Kerr nonlinear media could help unravel the complex dynamics of multivortex physics in quantum fluids.

\textit{Note:} We recently became aware of a similar but different approach on mapping the nonlinear optic wave equation to the Gross–Pitaevskii equation\cite{glorieuxParaxialFluidsLight2025}.

\begin{acknowledgments}
This work was supported by JSPS KAKENHI (Grant Nos. JP24H00424, JP23K03282, JP23K03305, JP22H05139, JP22H05131 and JP22H05132) and by JST FOREST Program (Grant Number JPMJFR236A, Japan).
\end{acknowledgments}

%\appendix
%\section{Appendixes}
%To start the appendixes, use the \verb+\appendix+ command.

%\nocite{*}

\bibliography{kelvinwave_opticalvortex}% Produces the bibliography via BibTeX.

\begin{thebibliography}{43}%
\makeatletter
\providecommand \@ifxundefined [1]{%
 \@ifx{#1\undefined}
}%
\providecommand \@ifnum [1]{%
 \ifnum #1\expandafter \@firstoftwo
 \else \expandafter \@secondoftwo
 \fi
}%
\providecommand \@ifx [1]{%
 \ifx #1\expandafter \@firstoftwo
 \else \expandafter \@secondoftwo
 \fi
}%
\providecommand \natexlab [1]{#1}%
\providecommand \enquote  [1]{``#1''}%
\providecommand \bibnamefont  [1]{#1}%
\providecommand \bibfnamefont [1]{#1}%
\providecommand \citenamefont [1]{#1}%
\providecommand \href@noop [0]{\@secondoftwo}%
\providecommand \href [0]{\begingroup \@sanitize@url \@href}%
\providecommand \@href[1]{\@@startlink{#1}\@@href}%
\providecommand \@@href[1]{\endgroup#1\@@endlink}%
\providecommand \@sanitize@url [0]{\catcode `\\12\catcode `\$12\catcode `\&12\catcode `\#12\catcode `\^12\catcode `\_12\catcode `\%12\relax}%
\providecommand \@@startlink[1]{}%
\providecommand \@@endlink[0]{}%
\providecommand \url  [0]{\begingroup\@sanitize@url \@url }%
\providecommand \@url [1]{\endgroup\@href {#1}{\urlprefix }}%
\providecommand \urlprefix  [0]{URL }%
\providecommand \Eprint [0]{\href }%
\providecommand \doibase [0]{https://doi.org/}%
\providecommand \selectlanguage [0]{\@gobble}%
\providecommand \bibinfo  [0]{\@secondoftwo}%
\providecommand \bibfield  [0]{\@secondoftwo}%
\providecommand \translation [1]{[#1]}%
\providecommand \BibitemOpen [0]{}%
\providecommand \bibitemStop [0]{}%
\providecommand \bibitemNoStop [0]{.\EOS\space}%
\providecommand \EOS [0]{\spacefactor3000\relax}%
\providecommand \BibitemShut  [1]{\csname bibitem#1\endcsname}%
\let\auto@bib@innerbib\@empty
%</preamble>
\bibitem [{\citenamefont {Allen}\ \emph {et~al.}(1992)\citenamefont {Allen}, \citenamefont {Beijersbergen}, \citenamefont {Spreeuw},\ and\ \citenamefont {Woerdman}}]{allenOrbitalAngularMomentum1992}%
  \BibitemOpen
  \bibfield  {author} {\bibinfo {author} {\bibfnamefont {L.}~\bibnamefont {Allen}}, \bibinfo {author} {\bibfnamefont {M.~W.}\ \bibnamefont {Beijersbergen}}, \bibinfo {author} {\bibfnamefont {R.~J.~C.}\ \bibnamefont {Spreeuw}},\ and\ \bibinfo {author} {\bibfnamefont {J.~P.}\ \bibnamefont {Woerdman}},\ }\bibfield  {title} {\bibinfo {title} {Orbital angular momentum of light and the transformation of {{Laguerre-Gaussian}} laser modes},\ }\href {https://doi.org/10.1103/PhysRevA.45.8185} {\bibfield  {journal} {\bibinfo  {journal} {Physical Review A}\ }\textbf {\bibinfo {volume} {45}},\ \bibinfo {pages} {8185} (\bibinfo {year} {1992})}\BibitemShut {NoStop}%
\bibitem [{\citenamefont {Wong}\ \emph {et~al.}(2012)\citenamefont {Wong}, \citenamefont {Kang}, \citenamefont {Lee}, \citenamefont {Biancalana}, \citenamefont {Conti}, \citenamefont {Weiss},\ and\ \citenamefont {Russell}}]{wongExcitationOrbitalAngular2012}%
  \BibitemOpen
  \bibfield  {author} {\bibinfo {author} {\bibfnamefont {G.~K.~L.}\ \bibnamefont {Wong}}, \bibinfo {author} {\bibfnamefont {M.~S.}\ \bibnamefont {Kang}}, \bibinfo {author} {\bibfnamefont {H.~W.}\ \bibnamefont {Lee}}, \bibinfo {author} {\bibfnamefont {F.}~\bibnamefont {Biancalana}}, \bibinfo {author} {\bibfnamefont {C.}~\bibnamefont {Conti}}, \bibinfo {author} {\bibfnamefont {T.}~\bibnamefont {Weiss}},\ and\ \bibinfo {author} {\bibfnamefont {P.~{\relax St}.~J.}\ \bibnamefont {Russell}},\ }\bibfield  {title} {\bibinfo {title} {Excitation of {{Orbital Angular Momentum Resonances}} in {{Helically Twisted Photonic Crystal Fiber}}},\ }\href {https://doi.org/10.1126/science.1223824} {\bibfield  {journal} {\bibinfo  {journal} {Science}\ }\textbf {\bibinfo {volume} {337}},\ \bibinfo {pages} {446} (\bibinfo {year} {2012})}\BibitemShut {NoStop}%
\bibitem [{\citenamefont {Hashiyada}\ and\ \citenamefont {Tanaka}(2024)}]{hashiyadaRapidModulationLeft2024}%
  \BibitemOpen
  \bibfield  {author} {\bibinfo {author} {\bibfnamefont {S.}~\bibnamefont {Hashiyada}}\ and\ \bibinfo {author} {\bibfnamefont {Y.~Y.}\ \bibnamefont {Tanaka}},\ }\bibfield  {title} {\bibinfo {title} {Rapid modulation of left- and right-handed optical vortices for precise measurements of helical dichroism},\ }\href {https://doi.org/10.1063/5.0203715} {\bibfield  {journal} {\bibinfo  {journal} {Review of Scientific Instruments}\ }\textbf {\bibinfo {volume} {95}},\ \bibinfo {pages} {053101} (\bibinfo {year} {2024})}\BibitemShut {NoStop}%
\bibitem [{\citenamefont {Taguchi}\ \emph {et~al.}(2025)\citenamefont {Taguchi}, \citenamefont {Fukui},\ and\ \citenamefont {Sasaki}}]{taguchiNanoscaleChiralityEnhancement2025}%
  \BibitemOpen
  \bibfield  {author} {\bibinfo {author} {\bibfnamefont {A.}~\bibnamefont {Taguchi}}, \bibinfo {author} {\bibfnamefont {Y.}~\bibnamefont {Fukui}},\ and\ \bibinfo {author} {\bibfnamefont {K.}~\bibnamefont {Sasaki}},\ }\bibfield  {title} {\bibinfo {title} {Nanoscale chirality enhancement using topology-designed three-dimensional dielectric nanogap antennas},\ }\href {https://doi.org/10.1103/PhysRevApplied.23.L021002} {\bibfield  {journal} {\bibinfo  {journal} {Physical Review Applied}\ }\textbf {\bibinfo {volume} {23}},\ \bibinfo {pages} {L021002} (\bibinfo {year} {2025})}\BibitemShut {NoStop}%
\bibitem [{\citenamefont {He}\ \emph {et~al.}(1995)\citenamefont {He}, \citenamefont {Friese}, \citenamefont {Heckenberg},\ and\ \citenamefont {{Rubinsztein-Dunlop}}}]{heDirectObservationTransfer1995}%
  \BibitemOpen
  \bibfield  {author} {\bibinfo {author} {\bibfnamefont {H.}~\bibnamefont {He}}, \bibinfo {author} {\bibfnamefont {M.~E.~J.}\ \bibnamefont {Friese}}, \bibinfo {author} {\bibfnamefont {N.~R.}\ \bibnamefont {Heckenberg}},\ and\ \bibinfo {author} {\bibfnamefont {H.}~\bibnamefont {{Rubinsztein-Dunlop}}},\ }\bibfield  {title} {\bibinfo {title} {Direct {{Observation}} of {{Transfer}} of {{Angular Momentum}} to {{Absorptive Particles}} from a {{Laser Beam}} with a {{Phase Singularity}}},\ }\href {https://doi.org/10.1103/PhysRevLett.75.826} {\bibfield  {journal} {\bibinfo  {journal} {Physical Review Letters}\ }\textbf {\bibinfo {volume} {75}},\ \bibinfo {pages} {826} (\bibinfo {year} {1995})}\BibitemShut {NoStop}%
\bibitem [{\citenamefont {Simpson}\ \emph {et~al.}(1997)\citenamefont {Simpson}, \citenamefont {Dholakia}, \citenamefont {Allen},\ and\ \citenamefont {Padgett}}]{simpsonMechanicalEquivalenceSpin1997}%
  \BibitemOpen
  \bibfield  {author} {\bibinfo {author} {\bibfnamefont {N.~B.}\ \bibnamefont {Simpson}}, \bibinfo {author} {\bibfnamefont {K.}~\bibnamefont {Dholakia}}, \bibinfo {author} {\bibfnamefont {L.}~\bibnamefont {Allen}},\ and\ \bibinfo {author} {\bibfnamefont {M.~J.}\ \bibnamefont {Padgett}},\ }\bibfield  {title} {\bibinfo {title} {Mechanical equivalence of spin and orbital angular momentum of light: An optical spanner},\ }\href {https://doi.org/10.1364/OL.22.000052} {\bibfield  {journal} {\bibinfo  {journal} {Optics Letters}\ }\textbf {\bibinfo {volume} {22}},\ \bibinfo {pages} {52} (\bibinfo {year} {1997})}\BibitemShut {NoStop}%
\bibitem [{\citenamefont {Toyoda}\ \emph {et~al.}(2013)\citenamefont {Toyoda}, \citenamefont {Takahashi}, \citenamefont {Takizawa}, \citenamefont {Tokizane}, \citenamefont {Miyamoto}, \citenamefont {Morita},\ and\ \citenamefont {Omatsu}}]{toyodaTransferLightHelicity2013}%
  \BibitemOpen
  \bibfield  {author} {\bibinfo {author} {\bibfnamefont {K.}~\bibnamefont {Toyoda}}, \bibinfo {author} {\bibfnamefont {F.}~\bibnamefont {Takahashi}}, \bibinfo {author} {\bibfnamefont {S.}~\bibnamefont {Takizawa}}, \bibinfo {author} {\bibfnamefont {Y.}~\bibnamefont {Tokizane}}, \bibinfo {author} {\bibfnamefont {K.}~\bibnamefont {Miyamoto}}, \bibinfo {author} {\bibfnamefont {R.}~\bibnamefont {Morita}},\ and\ \bibinfo {author} {\bibfnamefont {T.}~\bibnamefont {Omatsu}},\ }\bibfield  {title} {\bibinfo {title} {Transfer of {{Light Helicity}} to {{Nanostructures}}},\ }\href {https://doi.org/10.1103/PhysRevLett.110.143603} {\bibfield  {journal} {\bibinfo  {journal} {Physical Review Letters}\ }\textbf {\bibinfo {volume} {110}},\ \bibinfo {pages} {143603} (\bibinfo {year} {2013})}\BibitemShut {NoStop}%
\bibitem [{\citenamefont {Yuyama}\ \emph {et~al.}(2023)\citenamefont {Yuyama}, \citenamefont {Kawaguchi}, \citenamefont {Wei},\ and\ \citenamefont {Omatsu}}]{yuyamaFabricationArrayHemispherical2023}%
  \BibitemOpen
  \bibfield  {author} {\bibinfo {author} {\bibfnamefont {K.}~\bibnamefont {Yuyama}}, \bibinfo {author} {\bibfnamefont {H.}~\bibnamefont {Kawaguchi}}, \bibinfo {author} {\bibfnamefont {R.}~\bibnamefont {Wei}},\ and\ \bibinfo {author} {\bibfnamefont {T.}~\bibnamefont {Omatsu}},\ }\bibfield  {title} {\bibinfo {title} {Fabrication of an {{Array}} of {{Hemispherical Microlasers Using Optical Vortex Laser-Induced Forward Transfer}}},\ }\href {https://doi.org/10.1021/acsphotonics.3c01005} {\bibfield  {journal} {\bibinfo  {journal} {ACS Photonics}\ }\textbf {\bibinfo {volume} {10}},\ \bibinfo {pages} {4045} (\bibinfo {year} {2023})}\BibitemShut {NoStop}%
\bibitem [{\citenamefont {Matsumoto}\ \emph {et~al.}(2025)\citenamefont {Matsumoto}, \citenamefont {Masui},\ and\ \citenamefont {Hosokawa}}]{matsumotoHelicalSurfaceRelief2025}%
  \BibitemOpen
  \bibfield  {author} {\bibinfo {author} {\bibfnamefont {Y.}~\bibnamefont {Matsumoto}}, \bibinfo {author} {\bibfnamefont {K.}~\bibnamefont {Masui}},\ and\ \bibinfo {author} {\bibfnamefont {C.}~\bibnamefont {Hosokawa}},\ }\bibfield  {title} {\bibinfo {title} {Helical {{Surface Relief Formation}} by {{Two-Photon Polymerization Reaction Using}} a {{Femtosecond Optical Vortex Beam}}},\ }\href {https://doi.org/10.1021/acs.jpclett.4c03055} {\bibfield  {journal} {\bibinfo  {journal} {The Journal of Physical Chemistry Letters}\ }\textbf {\bibinfo {volume} {16}},\ \bibinfo {pages} {415} (\bibinfo {year} {2025})}\BibitemShut {NoStop}%
\bibitem [{\citenamefont {Toyoda}\ \emph {et~al.}(2023)\citenamefont {Toyoda}, \citenamefont {Su}, \citenamefont {Miyamoto}, \citenamefont {Sugiyama},\ and\ \citenamefont {Omatsu}}]{toyodaChiralCrystallizationManipulated2023}%
  \BibitemOpen
  \bibfield  {author} {\bibinfo {author} {\bibfnamefont {K.}~\bibnamefont {Toyoda}}, \bibinfo {author} {\bibfnamefont {H.-T.}\ \bibnamefont {Su}}, \bibinfo {author} {\bibfnamefont {K.}~\bibnamefont {Miyamoto}}, \bibinfo {author} {\bibfnamefont {T.}~\bibnamefont {Sugiyama}},\ and\ \bibinfo {author} {\bibfnamefont {T.}~\bibnamefont {Omatsu}},\ }\bibfield  {title} {\bibinfo {title} {Chiral crystallization manipulated by orbital angular momentum of light},\ }\href {https://doi.org/10.1364/OPTICA.478042} {\bibfield  {journal} {\bibinfo  {journal} {Optica}\ }\textbf {\bibinfo {volume} {10}},\ \bibinfo {pages} {332} (\bibinfo {year} {2023})}\BibitemShut {NoStop}%
\bibitem [{\citenamefont {Kivshar}\ \emph {et~al.}(1998)\citenamefont {Kivshar}, \citenamefont {Christou}, \citenamefont {Tikhonenko}, \citenamefont {{Luther-Davies}},\ and\ \citenamefont {Pismen}}]{kivsharDynamicsOpticalVortex1998}%
  \BibitemOpen
  \bibfield  {author} {\bibinfo {author} {\bibfnamefont {Y.~S.}\ \bibnamefont {Kivshar}}, \bibinfo {author} {\bibfnamefont {J.}~\bibnamefont {Christou}}, \bibinfo {author} {\bibfnamefont {V.}~\bibnamefont {Tikhonenko}}, \bibinfo {author} {\bibfnamefont {B.}~\bibnamefont {{Luther-Davies}}},\ and\ \bibinfo {author} {\bibfnamefont {L.~M.}\ \bibnamefont {Pismen}},\ }\bibfield  {title} {\bibinfo {title} {Dynamics of optical vortex solitons},\ }\href {https://doi.org/10.1016/S0030-4018(98)00149-7} {\bibfield  {journal} {\bibinfo  {journal} {Optics Communications}\ }\textbf {\bibinfo {volume} {152}},\ \bibinfo {pages} {198} (\bibinfo {year} {1998})}\BibitemShut {NoStop}%
\bibitem [{\citenamefont {Kivshar}\ and\ \citenamefont {Agrawal}(2003)}]{kivsharOpticalSolitonsFibers2003}%
  \BibitemOpen
  \bibfield  {author} {\bibinfo {author} {\bibfnamefont {Y.~S.}\ \bibnamefont {Kivshar}}\ and\ \bibinfo {author} {\bibfnamefont {G.~P.}\ \bibnamefont {Agrawal}},\ }\href@noop {} {\emph {\bibinfo {title} {Optical {{Solitons}}: {{From Fibers}} to {{Photonic Crystals}}}}},\ \bibinfo {edition} {1st}\ ed.\ (\bibinfo  {publisher} {Academic Press},\ \bibinfo {address} {Amsterdam Boston},\ \bibinfo {year} {2003})\BibitemShut {NoStop}%
\bibitem [{\citenamefont {Pelinovsky}\ \emph {et~al.}(1995)\citenamefont {Pelinovsky}, \citenamefont {Stepanyants},\ and\ \citenamefont {Kivshar}}]{pelinovskySelffocusingPlaneDark1995}%
  \BibitemOpen
  \bibfield  {author} {\bibinfo {author} {\bibfnamefont {D.~E.}\ \bibnamefont {Pelinovsky}}, \bibinfo {author} {\bibfnamefont {Y.~A.}\ \bibnamefont {Stepanyants}},\ and\ \bibinfo {author} {\bibfnamefont {Y.~S.}\ \bibnamefont {Kivshar}},\ }\bibfield  {title} {\bibinfo {title} {Self-focusing of plane dark solitons in nonlinear defocusing media},\ }\href {https://doi.org/10.1103/PhysRevE.51.5016} {\bibfield  {journal} {\bibinfo  {journal} {Physical Review E}\ }\textbf {\bibinfo {volume} {51}},\ \bibinfo {pages} {5016} (\bibinfo {year} {1995})}\BibitemShut {NoStop}%
\bibitem [{\citenamefont {Swartzlander}\ and\ \citenamefont {Law}(1992)}]{swartzlanderOpticalVortexSolitons1992}%
  \BibitemOpen
  \bibfield  {author} {\bibinfo {author} {\bibfnamefont {G.~A.}\ \bibnamefont {Swartzlander}}\ and\ \bibinfo {author} {\bibfnamefont {C.~T.}\ \bibnamefont {Law}},\ }\bibfield  {title} {\bibinfo {title} {Optical vortex solitons observed in {{Kerr}} nonlinear media},\ }\href {https://doi.org/10.1103/PhysRevLett.69.2503} {\bibfield  {journal} {\bibinfo  {journal} {Physical Review Letters}\ }\textbf {\bibinfo {volume} {69}},\ \bibinfo {pages} {2503} (\bibinfo {year} {1992})}\BibitemShut {NoStop}%
\bibitem [{\citenamefont {Snyder}\ \emph {et~al.}(1992)\citenamefont {Snyder}, \citenamefont {Poladian},\ and\ \citenamefont {Mitchell}}]{snyderStableBlackSelfguided1992}%
  \BibitemOpen
  \bibfield  {author} {\bibinfo {author} {\bibfnamefont {A.~W.}\ \bibnamefont {Snyder}}, \bibinfo {author} {\bibfnamefont {L.}~\bibnamefont {Poladian}},\ and\ \bibinfo {author} {\bibfnamefont {D.~J.}\ \bibnamefont {Mitchell}},\ }\bibfield  {title} {\bibinfo {title} {Stable black self-guided beams of circular symmetry in a bulk {{Kerr}} medium},\ }\href {https://doi.org/10.1364/OL.17.000789} {\bibfield  {journal} {\bibinfo  {journal} {Optics Letters}\ }\textbf {\bibinfo {volume} {17}},\ \bibinfo {pages} {789} (\bibinfo {year} {1992})}\BibitemShut {NoStop}%
\bibitem [{\citenamefont {Pitaevskii}(1961)}]{pitaevskiiVortexLinesImperfect1961}%
  \BibitemOpen
  \bibfield  {author} {\bibinfo {author} {\bibfnamefont {L.~P.}\ \bibnamefont {Pitaevskii}},\ }\bibfield  {title} {\bibinfo {title} {Vortex lines in an imperfect {{Bose}} gas},\ }\href@noop {} {\bibfield  {journal} {\bibinfo  {journal} {Sov. Phys. JETP}\ }\textbf {\bibinfo {volume} {13}},\ \bibinfo {pages} {451} (\bibinfo {year} {1961})}\BibitemShut {NoStop}%
\bibitem [{\citenamefont {Rozas}\ \emph {et~al.}(1997)\citenamefont {Rozas}, \citenamefont {Law},\ and\ \citenamefont {Swartzlander}}]{rozasPropagationDynamicsOptical1997}%
  \BibitemOpen
  \bibfield  {author} {\bibinfo {author} {\bibfnamefont {D.}~\bibnamefont {Rozas}}, \bibinfo {author} {\bibfnamefont {C.~T.}\ \bibnamefont {Law}},\ and\ \bibinfo {author} {\bibfnamefont {G.~A.}\ \bibnamefont {Swartzlander}},\ }\bibfield  {title} {\bibinfo {title} {Propagation dynamics of optical vortices},\ }\href {https://doi.org/10.1364/JOSAB.14.003054} {\bibfield  {journal} {\bibinfo  {journal} {JOSA B}\ }\textbf {\bibinfo {volume} {14}},\ \bibinfo {pages} {3054} (\bibinfo {year} {1997})}\BibitemShut {NoStop}%
\bibitem [{\citenamefont {Boyd}(2008)}]{boydNonlinearOpticsThird2008}%
  \BibitemOpen
  \bibfield  {author} {\bibinfo {author} {\bibfnamefont {R.~W.}\ \bibnamefont {Boyd}},\ }\href@noop {} {\emph {\bibinfo {title} {Nonlinear {{Optics}}, {{Third Edition}}}}},\ \bibinfo {edition} {3rd}\ ed.\ (\bibinfo  {publisher} {Academic Press},\ \bibinfo {address} {Burlington, MA},\ \bibinfo {year} {2008})\BibitemShut {NoStop}%
\bibitem [{\citenamefont {Donnelly}(1991)}]{donnellyQuantizedVorticesHelium1991}%
  \BibitemOpen
  \bibfield  {author} {\bibinfo {author} {\bibfnamefont {R.~J.}\ \bibnamefont {Donnelly}},\ }\href@noop {} {\emph {\bibinfo {title} {Quantized {{Vortices}} in {{Helium II}}}}},\ \bibinfo {edition} {1st}\ ed.\ (\bibinfo  {publisher} {Cambridge University Press},\ \bibinfo {address} {Cambridge England ; New York},\ \bibinfo {year} {1991})\BibitemShut {NoStop}%
\bibitem [{\citenamefont {Desyatnikov}\ \emph {et~al.}(2005)\citenamefont {Desyatnikov}, \citenamefont {Kivshar},\ and\ \citenamefont {Torner}}]{desyatnikovChapter5Optical2005}%
  \BibitemOpen
  \bibfield  {author} {\bibinfo {author} {\bibfnamefont {A.~S.}\ \bibnamefont {Desyatnikov}}, \bibinfo {author} {\bibfnamefont {Y.~S.}\ \bibnamefont {Kivshar}},\ and\ \bibinfo {author} {\bibfnamefont {L.}~\bibnamefont {Torner}},\ }\bibfield  {title} {\bibinfo {title} {Chapter 5 - {{Optical}} vortices and vortex solitons},\ }in\ \href {https://doi.org/10.1016/S0079-6638(05)47006-7} {\emph {\bibinfo {booktitle} {Progress in {{Optics}}}}},\ Vol.~\bibinfo {volume} {47},\ \bibinfo {editor} {edited by\ \bibinfo {editor} {\bibfnamefont {E.}~\bibnamefont {Wolf}}}\ (\bibinfo  {publisher} {Elsevier},\ \bibinfo {year} {2005})\ pp.\ \bibinfo {pages} {291--391}\BibitemShut {NoStop}%
\bibitem [{\citenamefont {Barenghi}\ \emph {et~al.}(2023)\citenamefont {Barenghi}, \citenamefont {Skrbek},\ and\ \citenamefont {Sreenivasan}}]{barenghiQuantumTurbulence2023}%
  \BibitemOpen
  \bibfield  {author} {\bibinfo {author} {\bibfnamefont {C.~F.}\ \bibnamefont {Barenghi}}, \bibinfo {author} {\bibfnamefont {L.}~\bibnamefont {Skrbek}},\ and\ \bibinfo {author} {\bibfnamefont {K.~R.}\ \bibnamefont {Sreenivasan}},\ }\href {https://doi.org/10.1017/9781009345651} {\emph {\bibinfo {title} {Quantum {{Turbulence}}}}}\ (\bibinfo  {publisher} {Cambridge University Press},\ \bibinfo {address} {Cambridge},\ \bibinfo {year} {2023})\BibitemShut {NoStop}%
\bibitem [{\citenamefont {Minowa}\ \emph {et~al.}(2025)\citenamefont {Minowa}, \citenamefont {Yasui}, \citenamefont {Nakagawa}, \citenamefont {Inui}, \citenamefont {Tsubota},\ and\ \citenamefont {Ashida}}]{minowaDirectExcitationKelvin2025}%
  \BibitemOpen
  \bibfield  {author} {\bibinfo {author} {\bibfnamefont {Y.}~\bibnamefont {Minowa}}, \bibinfo {author} {\bibfnamefont {Y.}~\bibnamefont {Yasui}}, \bibinfo {author} {\bibfnamefont {T.}~\bibnamefont {Nakagawa}}, \bibinfo {author} {\bibfnamefont {S.}~\bibnamefont {Inui}}, \bibinfo {author} {\bibfnamefont {M.}~\bibnamefont {Tsubota}},\ and\ \bibinfo {author} {\bibfnamefont {M.}~\bibnamefont {Ashida}},\ }\bibfield  {title} {\bibinfo {title} {Direct excitation of {{Kelvin}} waves on quantized vortices},\ }\href {https://doi.org/10.1038/s41567-024-02720-9} {\bibfield  {journal} {\bibinfo  {journal} {Nature Physics}\ }\textbf {\bibinfo {volume} {21}},\ \bibinfo {pages} {233} (\bibinfo {year} {2025})}\BibitemShut {NoStop}%
\bibitem [{\citenamefont {Saffman}(1993)}]{saffmanVortexDynamics1993}%
  \BibitemOpen
  \bibfield  {author} {\bibinfo {author} {\bibfnamefont {P.~G.}\ \bibnamefont {Saffman}},\ }\href {https://doi.org/10.1017/CBO9780511624063} {\emph {\bibinfo {title} {Vortex {{Dynamics}}}}},\ Cambridge {{Monographs}} on {{Mechanics}}\ (\bibinfo  {publisher} {Cambridge University Press},\ \bibinfo {address} {Cambridge},\ \bibinfo {year} {1993})\BibitemShut {NoStop}%
\bibitem [{\citenamefont {Kivshar}\ and\ \citenamefont {Yang}(1994)}]{kivsharDarkSolitonsBackgrounds1994}%
  \BibitemOpen
  \bibfield  {author} {\bibinfo {author} {\bibfnamefont {Y.~S.}\ \bibnamefont {Kivshar}}\ and\ \bibinfo {author} {\bibfnamefont {X.}~\bibnamefont {Yang}},\ }\bibfield  {title} {\bibinfo {title} {Dark solitons on backgrounds of finite extent},\ }\href {https://doi.org/10.1016/0030-4018(94)90109-0} {\bibfield  {journal} {\bibinfo  {journal} {Optics Communications}\ }\textbf {\bibinfo {volume} {107}},\ \bibinfo {pages} {93} (\bibinfo {year} {1994})}\BibitemShut {NoStop}%
\bibitem [{\citenamefont {Sprangle}\ \emph {et~al.}(1990)\citenamefont {Sprangle}, \citenamefont {Esarey},\ and\ \citenamefont {Ting}}]{sprangleNonlinearInteractionIntense1990}%
  \BibitemOpen
  \bibfield  {author} {\bibinfo {author} {\bibfnamefont {P.}~\bibnamefont {Sprangle}}, \bibinfo {author} {\bibfnamefont {E.}~\bibnamefont {Esarey}},\ and\ \bibinfo {author} {\bibfnamefont {A.}~\bibnamefont {Ting}},\ }\bibfield  {title} {\bibinfo {title} {Nonlinear interaction of intense laser pulses in plasmas},\ }\href {https://doi.org/10.1103/PhysRevA.41.4463} {\bibfield  {journal} {\bibinfo  {journal} {Physical Review A}\ }\textbf {\bibinfo {volume} {41}},\ \bibinfo {pages} {4463} (\bibinfo {year} {1990})}\BibitemShut {NoStop}%
\bibitem [{\citenamefont {Fetter}\ and\ \citenamefont {Svidzinsky}(2001)}]{fetterVorticesTrappedDilute2001}%
  \BibitemOpen
  \bibfield  {author} {\bibinfo {author} {\bibfnamefont {A.~L.}\ \bibnamefont {Fetter}}\ and\ \bibinfo {author} {\bibfnamefont {A.~A.}\ \bibnamefont {Svidzinsky}},\ }\bibfield  {title} {\bibinfo {title} {Vortices in a trapped dilute {{Bose-Einstein}} condensate},\ }\href {https://doi.org/10.1088/0953-8984/13/12/201} {\bibfield  {journal} {\bibinfo  {journal} {Journal of Physics: Condensed Matter}\ }\textbf {\bibinfo {volume} {13}},\ \bibinfo {pages} {R135} (\bibinfo {year} {2001})}\BibitemShut {NoStop}%
\bibitem [{\citenamefont {Zee}(2010)}]{zeeQuantumFieldTheory2010}%
  \BibitemOpen
  \bibfield  {author} {\bibinfo {author} {\bibfnamefont {A.}~\bibnamefont {Zee}},\ }\href@noop {} {\emph {\bibinfo {title} {Quantum {{Field Theory}} in a {{Nutshell}}}}}\ (\bibinfo  {publisher} {Princeton Univ Pr},\ \bibinfo {address} {Princeton Oxford},\ \bibinfo {year} {2010})\BibitemShut {NoStop}%
\bibitem [{\citenamefont {Thomson}(1880)}]{thomsonVibrationsColumnarVortex1880}%
  \BibitemOpen
  \bibfield  {author} {\bibinfo {author} {\bibfnamefont {W.}~\bibnamefont {Thomson}},\ }\bibfield  {title} {\bibinfo {title} {Vibrations of a columnar vortex},\ }\href {https://doi.org/10.1080/14786448008626912} {\bibfield  {journal} {\bibinfo  {journal} {The London, Edinburgh, and Dublin Philosophical Magazine and Journal of Science}\ }\textbf {\bibinfo {volume} {10}},\ \bibinfo {pages} {155} (\bibinfo {year} {1880})}\BibitemShut {NoStop}%
\bibitem [{\citenamefont {Fonda}\ \emph {et~al.}(2014)\citenamefont {Fonda}, \citenamefont {Meichle}, \citenamefont {Ouellette}, \citenamefont {Hormoz},\ and\ \citenamefont {Lathrop}}]{fondaDirectObservationKelvin2014}%
  \BibitemOpen
  \bibfield  {author} {\bibinfo {author} {\bibfnamefont {E.}~\bibnamefont {Fonda}}, \bibinfo {author} {\bibfnamefont {D.~P.}\ \bibnamefont {Meichle}}, \bibinfo {author} {\bibfnamefont {N.~T.}\ \bibnamefont {Ouellette}}, \bibinfo {author} {\bibfnamefont {S.}~\bibnamefont {Hormoz}},\ and\ \bibinfo {author} {\bibfnamefont {D.~P.}\ \bibnamefont {Lathrop}},\ }\bibfield  {title} {\bibinfo {title} {Direct observation of {{Kelvin}} waves excited by quantized vortex reconnection},\ }\href {https://doi.org/10.1073/pnas.1312536110} {\bibfield  {journal} {\bibinfo  {journal} {Proceedings of the National Academy of Sciences}\ }\textbf {\bibinfo {volume} {111}},\ \bibinfo {pages} {4707} (\bibinfo {year} {2014})}\BibitemShut {NoStop}%
\bibitem [{\citenamefont {Vinen}\ \emph {et~al.}(2003)\citenamefont {Vinen}, \citenamefont {Tsubota},\ and\ \citenamefont {Mitani}}]{vinenKelvinWaveCascadeVortex2003}%
  \BibitemOpen
  \bibfield  {author} {\bibinfo {author} {\bibfnamefont {W.~F.}\ \bibnamefont {Vinen}}, \bibinfo {author} {\bibfnamefont {M.}~\bibnamefont {Tsubota}},\ and\ \bibinfo {author} {\bibfnamefont {A.}~\bibnamefont {Mitani}},\ }\bibfield  {title} {\bibinfo {title} {Kelvin-{{Wave Cascade}} on a {{Vortex}} in {{Superfluid 4He}} at a {{Very Low Temperature}}},\ }\href {https://doi.org/10.1103/PhysRevLett.91.135301} {\bibfield  {journal} {\bibinfo  {journal} {Physical Review Letters}\ }\textbf {\bibinfo {volume} {91}},\ \bibinfo {pages} {135301} (\bibinfo {year} {2003})}\BibitemShut {NoStop}%
\bibitem [{\citenamefont {Kivotides}\ \emph {et~al.}(2001)\citenamefont {Kivotides}, \citenamefont {Vassilicos}, \citenamefont {Samuels},\ and\ \citenamefont {Barenghi}}]{kivotidesKelvinWavesCascade2001}%
  \BibitemOpen
  \bibfield  {author} {\bibinfo {author} {\bibfnamefont {D.}~\bibnamefont {Kivotides}}, \bibinfo {author} {\bibfnamefont {J.~C.}\ \bibnamefont {Vassilicos}}, \bibinfo {author} {\bibfnamefont {D.~C.}\ \bibnamefont {Samuels}},\ and\ \bibinfo {author} {\bibfnamefont {C.~F.}\ \bibnamefont {Barenghi}},\ }\bibfield  {title} {\bibinfo {title} {Kelvin {{Waves Cascade}} in {{Superfluid Turbulence}}},\ }\href {https://doi.org/10.1103/PhysRevLett.86.3080} {\bibfield  {journal} {\bibinfo  {journal} {Physical Review Letters}\ }\textbf {\bibinfo {volume} {86}},\ \bibinfo {pages} {3080} (\bibinfo {year} {2001})}\BibitemShut {NoStop}%
\bibitem [{\citenamefont {Rowlands}(1973)}]{rowlandsVibrationsQuantizedVortex1973}%
  \BibitemOpen
  \bibfield  {author} {\bibinfo {author} {\bibfnamefont {G.}~\bibnamefont {Rowlands}},\ }\bibfield  {title} {\bibinfo {title} {Vibrations of a quantized vortex in a weakly interacting {{Bose}} fluid},\ }\href {https://doi.org/10.1088/0305-4470/6/3/007} {\bibfield  {journal} {\bibinfo  {journal} {Journal of Physics A: Mathematical, Nuclear and General}\ }\textbf {\bibinfo {volume} {6}},\ \bibinfo {pages} {322} (\bibinfo {year} {1973})}\BibitemShut {NoStop}%
\bibitem [{\citenamefont {Roberts}(2003)}]{robertsVortexWavesCompressible2003}%
  \BibitemOpen
  \bibfield  {author} {\bibinfo {author} {\bibfnamefont {P.~H.}\ \bibnamefont {Roberts}},\ }\bibfield  {title} {\bibinfo {title} {On vortex waves in compressible fluids. {{II}}. {{The}} condensate vortex},\ }\href {https://doi.org/10.1098/rspa.2002.1033} {\bibfield  {journal} {\bibinfo  {journal} {Proceedings of the Royal Society of London. Series A: Mathematical, Physical and Engineering Sciences}\ }\textbf {\bibinfo {volume} {459}},\ \bibinfo {pages} {597} (\bibinfo {year} {2003})}\BibitemShut {NoStop}%
\bibitem [{\citenamefont {Kobayashi}\ and\ \citenamefont {Nitta}(2014)}]{kobayashiKelvinModesNambu2014a}%
  \BibitemOpen
  \bibfield  {author} {\bibinfo {author} {\bibfnamefont {M.}~\bibnamefont {Kobayashi}}\ and\ \bibinfo {author} {\bibfnamefont {M.}~\bibnamefont {Nitta}},\ }\bibfield  {title} {\bibinfo {title} {Kelvin modes as {{Nambu}}--{{Goldstone}} modes along superfluid vortices and relativistic strings: {{Finite}} volume size effects},\ }\href {https://doi.org/10.1093/ptep/ptu017} {\bibfield  {journal} {\bibinfo  {journal} {Progress of Theoretical and Experimental Physics}\ }\textbf {\bibinfo {volume} {2014}},\ \bibinfo {pages} {021B01} (\bibinfo {year} {2014})}\BibitemShut {NoStop}%
\bibitem [{\citenamefont {Tang}\ \emph {et~al.}(2023)\citenamefont {Tang}, \citenamefont {Guo}, \citenamefont {Kobayashi}, \citenamefont {Yui}, \citenamefont {Tsubota},\ and\ \citenamefont {Kanai}}]{tangImagingQuantizedVortex2023}%
  \BibitemOpen
  \bibfield  {author} {\bibinfo {author} {\bibfnamefont {Y.}~\bibnamefont {Tang}}, \bibinfo {author} {\bibfnamefont {W.}~\bibnamefont {Guo}}, \bibinfo {author} {\bibfnamefont {H.}~\bibnamefont {Kobayashi}}, \bibinfo {author} {\bibfnamefont {S.}~\bibnamefont {Yui}}, \bibinfo {author} {\bibfnamefont {M.}~\bibnamefont {Tsubota}},\ and\ \bibinfo {author} {\bibfnamefont {T.}~\bibnamefont {Kanai}},\ }\bibfield  {title} {\bibinfo {title} {Imaging quantized vortex rings in superfluid helium to evaluate quantum dissipation},\ }\href {https://doi.org/10.1038/s41467-023-38787-w} {\bibfield  {journal} {\bibinfo  {journal} {Nature Communications}\ }\textbf {\bibinfo {volume} {14}},\ \bibinfo {pages} {2941} (\bibinfo {year} {2023})}\BibitemShut {NoStop}%
\bibitem [{\citenamefont {M{\"a}kinen}\ \emph {et~al.}(2023)\citenamefont {M{\"a}kinen}, \citenamefont {Autti}, \citenamefont {Heikkinen}, \citenamefont {Hosio}, \citenamefont {H{\"a}nninen}, \citenamefont {L'vov}, \citenamefont {Walmsley}, \citenamefont {Zavjalov},\ and\ \citenamefont {Eltsov}}]{makinenRotatingQuantumWave2023}%
  \BibitemOpen
  \bibfield  {author} {\bibinfo {author} {\bibfnamefont {J.~T.}\ \bibnamefont {M{\"a}kinen}}, \bibinfo {author} {\bibfnamefont {S.}~\bibnamefont {Autti}}, \bibinfo {author} {\bibfnamefont {P.~J.}\ \bibnamefont {Heikkinen}}, \bibinfo {author} {\bibfnamefont {J.~J.}\ \bibnamefont {Hosio}}, \bibinfo {author} {\bibfnamefont {R.}~\bibnamefont {H{\"a}nninen}}, \bibinfo {author} {\bibfnamefont {V.~S.}\ \bibnamefont {L'vov}}, \bibinfo {author} {\bibfnamefont {P.~M.}\ \bibnamefont {Walmsley}}, \bibinfo {author} {\bibfnamefont {V.~V.}\ \bibnamefont {Zavjalov}},\ and\ \bibinfo {author} {\bibfnamefont {V.~B.}\ \bibnamefont {Eltsov}},\ }\bibfield  {title} {\bibinfo {title} {Rotating quantum wave turbulence},\ }\href {https://doi.org/10.1038/s41567-023-01966-z} {\bibfield  {journal} {\bibinfo  {journal} {Nature Physics}\ }\textbf {\bibinfo {volume} {19}},\ \bibinfo {pages} {898} (\bibinfo {year} {2023})}\BibitemShut {NoStop}%
\bibitem [{\citenamefont {Peretti}\ \emph {et~al.}(2023)\citenamefont {Peretti}, \citenamefont {Vessaire}, \citenamefont {Durozoy},\ and\ \citenamefont {Gibert}}]{perettiDirectVisualizationQuantum2023}%
  \BibitemOpen
  \bibfield  {author} {\bibinfo {author} {\bibfnamefont {C.}~\bibnamefont {Peretti}}, \bibinfo {author} {\bibfnamefont {J.}~\bibnamefont {Vessaire}}, \bibinfo {author} {\bibfnamefont {E.}~\bibnamefont {Durozoy}},\ and\ \bibinfo {author} {\bibfnamefont {M.}~\bibnamefont {Gibert}},\ }\bibfield  {title} {\bibinfo {title} {Direct visualization of the quantum vortex lattice structure, oscillations, and destabilization in rotating {{4He}}},\ }\href {https://doi.org/10.1126/sciadv.adh2899} {\bibfield  {journal} {\bibinfo  {journal} {Science Advances}\ }\textbf {\bibinfo {volume} {9}},\ \bibinfo {pages} {eadh2899} (\bibinfo {year} {2023})}\BibitemShut {NoStop}%
\bibitem [{\citenamefont {Minowa}\ \emph {et~al.}(2022)\citenamefont {Minowa}, \citenamefont {Aoyagi}, \citenamefont {Inui}, \citenamefont {Nakagawa}, \citenamefont {Asaka}, \citenamefont {Tsubota},\ and\ \citenamefont {Ashida}}]{minowaVisualizationQuantizedVortex2022}%
  \BibitemOpen
  \bibfield  {author} {\bibinfo {author} {\bibfnamefont {Y.}~\bibnamefont {Minowa}}, \bibinfo {author} {\bibfnamefont {S.}~\bibnamefont {Aoyagi}}, \bibinfo {author} {\bibfnamefont {S.}~\bibnamefont {Inui}}, \bibinfo {author} {\bibfnamefont {T.}~\bibnamefont {Nakagawa}}, \bibinfo {author} {\bibfnamefont {G.}~\bibnamefont {Asaka}}, \bibinfo {author} {\bibfnamefont {M.}~\bibnamefont {Tsubota}},\ and\ \bibinfo {author} {\bibfnamefont {M.}~\bibnamefont {Ashida}},\ }\bibfield  {title} {\bibinfo {title} {Visualization of quantized vortex reconnection enabled by laser ablation},\ }\href {https://doi.org/10.1126/sciadv.abn1143} {\bibfield  {journal} {\bibinfo  {journal} {Science Advances}\ }\textbf {\bibinfo {volume} {8}},\ \bibinfo {pages} {eabn1143} (\bibinfo {year} {2022})}\BibitemShut {NoStop}%
\bibitem [{\citenamefont {Pethick}\ and\ \citenamefont {Smith}(2008)}]{pethickBoseEinsteinCondensation2008}%
  \BibitemOpen
  \bibfield  {author} {\bibinfo {author} {\bibfnamefont {C.~J.}\ \bibnamefont {Pethick}}\ and\ \bibinfo {author} {\bibfnamefont {H.}~\bibnamefont {Smith}},\ }\href {https://doi.org/10.1017/CBO9780511802850} {\emph {\bibinfo {title} {Bose--{{Einstein Condensation}} in {{Dilute Gases}}}}},\ \bibinfo {edition} {2nd}\ ed.\ (\bibinfo  {publisher} {Cambridge University Press},\ \bibinfo {address} {Cambridge},\ \bibinfo {year} {2008})\BibitemShut {NoStop}%
\bibitem [{\citenamefont {Ishimori}(1986)}]{ishimoriDynamicsTopologicalVortices1986}%
  \BibitemOpen
  \bibfield  {author} {\bibinfo {author} {\bibfnamefont {Y.}~\bibnamefont {Ishimori}},\ }\bibfield  {title} {\bibinfo {title} {Dynamics of {{Topological Vortices}} in {{Two-Dimensional Nonlinear Wave Systems}}. {{I}}. {{Lagrangian Approach}}},\ }\href {https://doi.org/10.1143/JPSJ.55.82} {\bibfield  {journal} {\bibinfo  {journal} {Journal of the Physical Society of Japan}\ }\textbf {\bibinfo {volume} {55}},\ \bibinfo {pages} {82} (\bibinfo {year} {1986})}\BibitemShut {NoStop}%
\bibitem [{\citenamefont {Wang}\ \emph {et~al.}(2020)\citenamefont {Wang}, \citenamefont {Yuan}, \citenamefont {Wang}, \citenamefont {Xiao},\ and\ \citenamefont {Jia}}]{wangMeasurementKerrNonlinear2020}%
  \BibitemOpen
  \bibfield  {author} {\bibinfo {author} {\bibfnamefont {S.}~\bibnamefont {Wang}}, \bibinfo {author} {\bibfnamefont {J.}~\bibnamefont {Yuan}}, \bibinfo {author} {\bibfnamefont {L.}~\bibnamefont {Wang}}, \bibinfo {author} {\bibfnamefont {L.}~\bibnamefont {Xiao}},\ and\ \bibinfo {author} {\bibfnamefont {S.}~\bibnamefont {Jia}},\ }\bibfield  {title} {\bibinfo {title} {Measurement of the {{Kerr}} nonlinear refractive index of the {{Rb}} vapor based on an optical frequency comb using the z-scan method},\ }\href {https://doi.org/10.1364/OE.413350} {\bibfield  {journal} {\bibinfo  {journal} {Optics Express}\ }\textbf {\bibinfo {volume} {28}},\ \bibinfo {pages} {38334} (\bibinfo {year} {2020})}\BibitemShut {NoStop}%
\bibitem [{\citenamefont {Bozinovic}\ \emph {et~al.}(2013)\citenamefont {Bozinovic}, \citenamefont {Yue}, \citenamefont {Ren}, \citenamefont {Tur}, \citenamefont {Kristensen}, \citenamefont {Huang}, \citenamefont {Willner},\ and\ \citenamefont {Ramachandran}}]{bozinovicTerabitScaleOrbitalAngular2013}%
  \BibitemOpen
  \bibfield  {author} {\bibinfo {author} {\bibfnamefont {N.}~\bibnamefont {Bozinovic}}, \bibinfo {author} {\bibfnamefont {Y.}~\bibnamefont {Yue}}, \bibinfo {author} {\bibfnamefont {Y.}~\bibnamefont {Ren}}, \bibinfo {author} {\bibfnamefont {M.}~\bibnamefont {Tur}}, \bibinfo {author} {\bibfnamefont {P.}~\bibnamefont {Kristensen}}, \bibinfo {author} {\bibfnamefont {H.}~\bibnamefont {Huang}}, \bibinfo {author} {\bibfnamefont {A.~E.}\ \bibnamefont {Willner}},\ and\ \bibinfo {author} {\bibfnamefont {S.}~\bibnamefont {Ramachandran}},\ }\bibfield  {title} {\bibinfo {title} {Terabit-{{Scale Orbital Angular Momentum Mode Division Multiplexing}} in {{Fibers}}},\ }\href {https://doi.org/10.1126/science.1237861} {\bibfield  {journal} {\bibinfo  {journal} {Science}\ }\textbf {\bibinfo {volume} {340}},\ \bibinfo {pages} {1545} (\bibinfo {year} {2013})}\BibitemShut {NoStop}%
\bibitem [{\citenamefont {Glorieux}\ \emph {et~al.}(2025)\citenamefont {Glorieux}, \citenamefont {Piekarski}, \citenamefont {Schibler}, \citenamefont {Aladjidi},\ and\ \citenamefont {{Baker-Rasooli}}}]{glorieuxParaxialFluidsLight2025}%
  \BibitemOpen
  \bibfield  {author} {\bibinfo {author} {\bibfnamefont {Q.}~\bibnamefont {Glorieux}}, \bibinfo {author} {\bibfnamefont {C.}~\bibnamefont {Piekarski}}, \bibinfo {author} {\bibfnamefont {Q.}~\bibnamefont {Schibler}}, \bibinfo {author} {\bibfnamefont {T.}~\bibnamefont {Aladjidi}},\ and\ \bibinfo {author} {\bibfnamefont {M.}~\bibnamefont {{Baker-Rasooli}}},\ }\href {https://doi.org/10.48550/arXiv.2504.06262} {\bibinfo {title} {Paraxial fluids of light}} (\bibinfo {year} {2025}),\ \Eprint {https://arxiv.org/abs/2504.06262} {arXiv:2504.06262 [cond-mat]} \BibitemShut {NoStop}%
\end{thebibliography}%

\end{document}